  \documentclass[fleqn]{2022AMSE-arxiv}

\usepackage{bm}
\newcommand\Rey{\mbox{\textit{Re}}}  
\newcommand{\comm}[1]{}
\newcommand{\q}[1]{\textquotedblleft #1\textquotedblright}

\begin{document}




\title{Wake instability of a fixed spherical droplet with a high drop-to-fluid viscosity ratio}

	

\author[1,2]{Pengyu Shi}{p.shi@hzdr.de}%
\author[1]{\'{E}ric Climent}{eric.climent@imft.fr}
\author[1]{Dominique Legendre}{dominique.legendre@imft.fr}

\AuthorMark{P. Shi}

\address[1]{IMFT (Institut de M\'ecanique des Fluides de Toulouse), Universit\'e de Toulouse; CNRS, INPT, UPS; F-31400 Toulouse, France}
\address[2]{Helmholtz-Zentrum Dresden – Rossendorf, Institute of Fluid Dynamics, 01328 Dresden, Germany}


\abstract{Direct numerical simulations of a uniform flow past a fixed spherical droplet are performed to investigate the parameter range within which the axisymmetric flow becomes unstable due to an external flow bifurcation. The hydrodynamics is governed by three dimensionless numbers: the viscosity ratio, $\mu^\ast$, and the external and internal Reynolds numbers, $\Rey^e$ and $\Rey^i$, respectively. The drop-to-fluid density ratio is related to these parameters as $\rho^\ast=\mu^\ast \Rey^i/\Rey^e$. This study focuses on highly viscous droplets with $\mu^\ast \geq 5$, where wake instability is driven by the vorticity flux transferred from the droplet surface into the surrounding fluid. By analysing the wake structure, we confirm that the onset of the external bifurcation is linked to the tilting of the azimuthal vorticity, $\omega_\phi$, in the wake and that the bifurcation occurs once the isocontours of $\omega_\phi$ align nearly perpendicular to the symmetry axis. We propose an empirical criterion for predicting the onset of the external bifurcation, formulated in terms of the maximum vorticity on the external side of the droplet surface. This criterion is applicable for sufficiently high $\Rey^i$ and holds over a wide range of $\mu^\ast$ and $\Rey^e$. Additionally, we examine the bifurcation sequence for two specific external Reynolds numbers, $\Rey^e=300$ and $\Rey^e=500$, and show that, beyond a critical viscosity ratio, the axisymmetric wake first transitions to a steady planar-symmetric state before undergoing a secondary Hopf bifurcation. Finally, we highlight the influence of $\Rey^i$ on external bifurcation and show that, at moderate $\Rey^i$, wake instability may set in at a lower vorticity threshold than predicted by our criterion. These findings provide new insights into the external flow bifurcation of viscous droplets.}


\setlength{\textheight}{23.6cm}
\thispagestyle{empty}

\maketitle
\setlength{\parindent}{1em}

\vspace{-1mm}
\begin{multicols}{2}
\section{Introduction} \label{sec:int}
Bubbles, droplets, and particles with axisymmetric shapes may follow an unstable path when rising or falling freely in a fluid at rest under the influence of buoyancy and gravity \cite{2012_Ern}.  Understanding the mechanism of path instability is of fundamental interest and crucial for applications such as mechanical and chemical processes involving multiphase flows \cite{2000_Magnaudet, mathai2020bubbly, bourouiba2021fluid, legendre2024fluid}. It is generally accepted that the path instabilities of axisymmetric bodies is closely related to wake instability when such bodies are fixed in a uniform flow. Hence, a first step in understanding path instability is to determine the parameter conditions under which the wake first becomes unstable. For bubbles and particles, numerous studies exist, and the knowledge on wake instability is well-established \cite{ghidersa2000breaking, mougin2002path, 2007_Magnaudet}. However, for droplets, our understanding remains limited, which is the focus of this work.

For a uniform flow past a fixed droplet, the axisymmetry of the flow may break due to either an internal or an external flow bifurcation. The former typically occurs for droplets with a low-to-moderate viscosity ratio, and relevant studies have appeared in \cite{edelmann2017numerical, rachih2019etude, gode2024flow, 2024_Shi_drop}, and more recently, in our companion work \cite{2025_Shi_drop}. The external bifurcation, on the other hand, occurs for highly viscous droplets, in which case the surface vorticity flux transferred into the surrounding fluid is strong enough to trigger wake instability \cite{dandy1989buoyancy, blanco1995structure, 2007_Magnaudet}. While the underlying wake instability mechanism may be similar to that for particles and oblate spheroidal bubbles, a systematic investigation of the parameter conditions leading to external bifurcation for such droplets is still lacking. The present work aims to fill this gap. Specifically, we seek to:
\begin{enumerate}
    \item confirm the similarity between the external bifurcation mechanisms of droplets, oblate spheroidal bubbles, and particles, and
    \item propose an empirical criterion for determining the onset of the external bifurcation for highly viscous droplets that may be used within a wide parameter range of physical properties.
\end{enumerate}

\section{Problem statement and numerical approach} \label{sec:pro}
We consider a spherical droplet of radius $R$, density $\rho^i$, and dynamic viscosity $\mu^i$ that is fixed in a Newtonian fluid of density $\rho^e$ and dynamic viscosity $\mu^e$. The external flow far from the droplet is a uniform stream along $\boldsymbol{e}_x$, described by $\boldsymbol{u}^\infty = u_{rel}\boldsymbol{e}_x$, where $u_{rel}$ represents the slip velocity of the external fluid relative to the droplet. The entire flow field is governed by the incompressible Navier–Stokes equations. The boundary at the droplet surface satisfies a non-penetration condition, along with continuity of tangential velocity and shear stress. 
The steady-state solution of the problem is characterized by three dimensionless numbers: the viscosity ratio $\mu^\ast=\mu^i/\mu^e$, the external Reynolds number $\Rey^e$, and the internal Reynolds number $\Rey^i$. The latter two are defined as
\begin{equation}
\Rey^e=\frac{\rho^e u_{rel} (2R)}{\mu^e}, \quad
\Rey^i=\frac{\rho^i u_{rel} (2R)}{\mu^i}.
\label{eq:def_var}
\end{equation}
The drop-to-fluid density ratio can be expressed in terms of these three parameters as $\rho^\ast=\mu^\ast\,\Rey^i/\Rey^e$. 

The simulations were performed using the JADIM code developed at IMFT. This code has previously been applied to simulate three-dimensional flows around spherical bubbles and particles, as well as the associated hydrodynamic forces \cite{legendre1998lift, adoua2009reversal, shi2020hydrodynamic, shi2021drag} and has been recently extended to compute three-dimensional flows around and inside spherical droplets \cite{legendre2019basset, rachih2020numerical, gode2023basset, 2024_Shi_drop}. The reader is referred to \cite{2024_Shi_drop} for details regarding the numerical implementation, including convergence analysis with the mesh grid, boundary conditions, and validation tests confirming the reliability of the numerical approach. 

\section{Transition sequence} \label{tra_seq}
For the selected cases discussed, unless stated otherwise, the internal Reynolds number $\Rey^i$ is fixed at 1000 for $\Rey^e\leq500$ and reduced to 500 for larger $\Rey^e$. All simulations start from an initially axisymmetric flow and are carried out up to a physical time of $1000\,u_{rel}/R$. Hence, the \q{final} state achieved may not correspond to the behaviour for $t\,u_{rel}/R\to\infty$. For instance, in the series of cases with $(\Rey^e,\Rey^i)=(300,1000)$ to be discussed later, internal flow bifurcation may occur beyond $t\,u_{rel}/R=1000$ for $\mu^\ast$ up to approximately 12. These long-term behaviours, which are characteristic of internal bifurcation at moderate $\mu^\ast$, are not discussed here but are detailed in a companion work \cite{2025_Shi_drop}.

\subsection{Axisymmetric wake} \label{axi_wak}

\begin{figure*}[!htb]
\centerline{\includegraphics[scale=0.75]{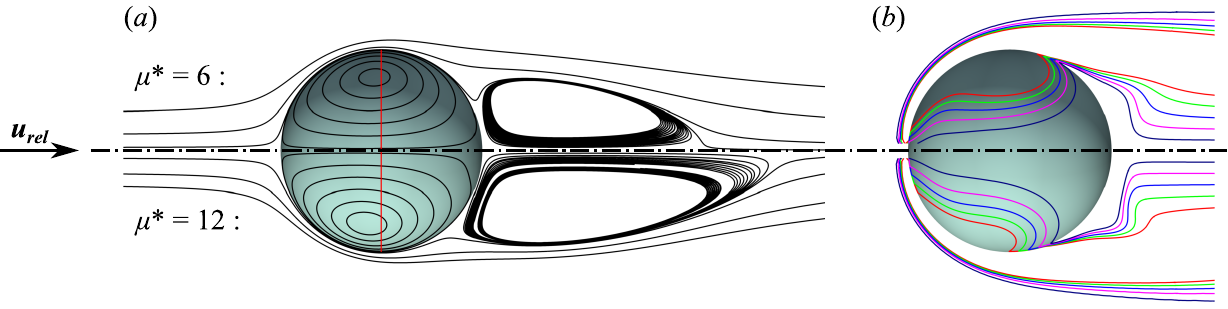}}
\caption{$(a)$ Streamlines and $(b)$ isocontours of the azimuthal vorticity $\omega_\phi$ around the droplet for $\mu^\ast = 6$ (top panels) and 12 (bottom panels) with $(\Rey^e,\Rey^i)=(300,1000)$. In $(a)$, the vertical red line denotes $x = 0$. In $(b)$, coloured lines represent $-\omega_\phi R/u_{rel} = 1$ (red), 0.8 (green), 0.6 (blue), 0.4 (magenta), and 0.2 (navy).}
\label{fig:axi_streamline}
\end{figure*}
We begin by revisiting the flow regime in which the wake remains axisymmetric. Prior studies \cite{feng2001drag, rachih2020numerical, gode2024flow} have shown that within this regime, a standing eddy forms in the wake when $\mu^\ast$ exceeds a critical threshold that depends on $\Rey^e$. According to \cite{rachih2020numerical} and \cite{gode2024flow}, this critical viscosity ratio is approximately 3 at $\Rey^e=200$ and becomes nearly independent of $\Rey^e$ at higher $\Rey^e$. Indeed, in all our simulations where $\Rey^e\geq200$, a standing eddy in the wake already appears at $\mu^\ast=5$, the smallest $\mu^\ast$ considered in our study. Figure \ref{fig:axi_streamline}$(a)$ shows the flow structure at $\Rey^e=300$ for $\mu^\ast=6$ (top) and $\mu^\ast=12$ (bottom). Inside the droplet, no flow separation occurs at the rear, but the flow exhibits a certain degree of fore-aft asymmetry, particularly at larger $\mu^\ast$, distinguishing it from a Hill's spherical vortex. Outside the droplet, the flow separates at the rear and reattaches along the symmetry axis at a distance that increases with $\mu^\ast$. Panel $(b)$ presents selected isocontours of the azimuthal vorticity $\omega_\phi$ around the two droplets. The most striking feature is that just behind the droplet, the isocontours tend to align almost perpendicularly to the symmetry axis, particularly at $\mu^\ast=12$. Prior work \cite{2007_Magnaudet} has concluded that such a situation is inherently unstable, as the streamwise gradient of $\omega_\phi$ and, consequently, the viscous term in the budget equation of $\omega_\phi$, become infinitely large when the $\omega_\phi$ isocontours turn perpendicular to the symmetry axis. Indeed, at a slightly larger viscosity ratio ($\mu^\ast=15$), axisymmetry breaks down, indicating that the underlying mechanism of symmetry breaking is essentially the same as that observed for particles and oblate spheroidal bubbles.

\subsection{The unstable $(\Rey^e,\,\mu^\ast)$ regime at high $\Rey^i$}
\label{uns_ran}
The breaking of axisymmetry can be tracked by examining the perturbation energy, for which a convenient measure is the mean energy of the azimuthal flow velocity component \cite{thompson2001kinematics, 2007_Magnaudet, 2025_Shi_drop}:
\begin{equation}
E^k=\frac{1}{\rho^e V_s u_{rel}^2} \int_{V^k}{\rho^k ||\boldsymbol{u}_\varphi^k||^2\mathrm{d}V^k},
\label{eq:energy}
\end{equation}
where $V_s=4 \pi R^3/3$ is the volume of the droplet, and $\boldsymbol{u}_\varphi^k$ is the azimuthal component of the local velocity. Here, $E^k$ (respectively $V^k$) with $k=i$ or $e$ denotes the azimuthal energy (respectively the domain) inside or outside the droplet. The mean azimuthal energy of the entire flow field is then given by $E= E^i +E^e$, which becomes positive as soon as the axisymmetry of the base flow breaks down.

\begin{figure*}[!htb]
\centerline{\includegraphics[scale=0.75]{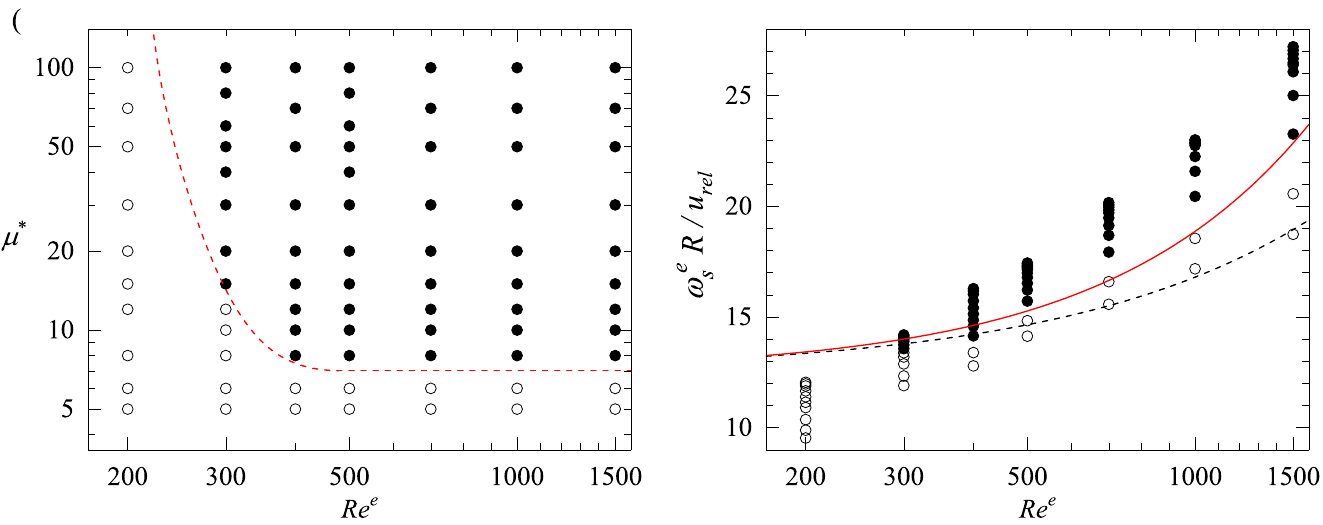}}
\caption{$(a)$ Phase diagram indicating the parameter range within which the axisymmetric flow becomes unstable. $(b)$ Maximum external surface vorticity as a function of $\Rey^e$. In both panels: $\circ$ denotes stable cases, while $\bullet$ denotes unstable cases. In $(a)$, the red dashed line approximates the boundary separating the unstable and stable regimes. In $(b)$, for each iso-$\Rey^e$ data series, $\mu^\ast$ increases from bottom to top; the dashed line represents the empirical correlation from \cite{2007_Magnaudet}, while the solid red line corresponds to Eq.\,\eqref{eq:cri_vorE}.}
\label{fig:vor_e}       
\end{figure*}

Within the parameter ranges of $(\Rey^e,\,\mu^\ast)$ explored in our work, the symmetry breaking is always driven by an external flow bifurcation. That is, for all cases where $E>0$, we found $E^i\approx0$ and hence $E\approx E^e$. \textcolor{black}{A typical example of this is provided in the next section, where figure \ref{fig:e_vs_t}$(a)$ compares the time evolutions of $E$ and $E^e$ for a droplet with $(\mu^\ast, \Rey^e, \Rey^i)=(20, 300, 1000)$.} The vanishingly small $E^i$ indicates that the internal flow remains axisymmetric, even though the external flow (and consequently the wake) becomes strongly asymmetric following the external bifurcation. Figure \ref{fig:vor_e}$(a)$ presents the phase diagram showing whether the entire axisymmetric flow is stable within the considered ranges of $(\Rey^e,\,\mu^\ast)$. The axisymmetric flow remains stable for $\Rey^e=200$ irrespective of $\mu^\ast$ but breaks down for $\Rey^e\geq300$ beyond a critical viscosity ratio. This critical $\mu^\ast$ is approximately 15 for $\Rey=300$ and decreases to about 8 for $\Rey\geq400$.

The mechanism governing the external flow bifurcation has been well described by \cite{2007_Magnaudet}. Specifically, at the threshold, the azimuthal mode with wavenumber $m=1$ first becomes unstable, resulting in a steady but asymmetric flow characterised by a pair of counter-rotating streamwise vortices in the wake \cite{ghidersa2000breaking, yang2007linear, tchoufag2013linear}. In \cite{2007_Magnaudet}, it was found that the primary wake instability occurs when the maximum vorticity generated on the external surface of the body, $\omega_s^e=\max{||\boldsymbol{\omega}_S^e||}$, exceeds a critical value that increases with $\Rey^e$. The external surface vorticity $||\boldsymbol{\omega}_S^e||$ is defined as
\begin{equation}
\boldsymbol\omega_S^e = \lim_{\;\: r \rightarrow R^+}\boldsymbol\omega^e - \left( \boldsymbol\omega^e \cdot \boldsymbol{n}\right) \boldsymbol{n},
\label{eq:surf_vor}
\end{equation}
where $\boldsymbol\omega^e = \nabla \times \boldsymbol{u}^e$ is the external vorticity, and $\boldsymbol{n}$ is the outward unit normal to the droplet surface.

Figure \ref{fig:vor_e}$(b)$ summarises the resulting $\omega_s^e$ (normalised with $u_{rel}/R$) at different $\Rey^e$. For each iso-$\Rey^e$ data series, $\omega_s^e$ increases with increasing $\mu^\ast$ ($\mu^\ast$ increases from bottom to top). The critical $\omega_s^e$ is approximately $13.5\,u_{rel}/R$ for $\Rey^e=300$ and increases to a value in the range $(18.5,\,20.5]$ for $\Rey^e=1000$, in line with the trend reported by \cite{2007_Magnaudet}. Also shown in panel $(b)$ is the empirical correlation for the critical $\omega_s^e$ from this prior work (dashed line). \textcolor{black}{It should be noted that in \cite{2007_Magnaudet}, the critical $\omega_s^e$ at a fixed $\Rey^e$ was reached by increasing the aspect ratio of an oblate spheroidal bubble, whereas in the present study, it is reached by increasing the viscosity ratio $\mu^\ast$ of a spherical droplet.} The agreement with our DNS results is satisfactory for $\Rey^e$ up to 500, beyond which it underestimates the critical $\omega_s^e$ (and consequently the critical $\mu^\ast$). Following the approach of \cite{2007_Magnaudet}, we find that the critical $\omega_s^e$ marking the onset of primary instability in a droplet wake can be better captured using (solid red line in panel $(b)$):
\begin{equation}
\omega_c^e\, R/u_{rel} \approx 12.5+1.6 \times 10^{-3} \left( \Rey^e \right)^{1.2}.
\label{eq:cri_vorE}
\end{equation}
Specifically, in the solid-sphere limit $\mu^\ast\to\infty$, $\omega_s^e\, R/u_{rel}$ varies with $\Rey^e$ as $1.5\left(1 - 0.12\sqrt{\Rey^e} + 0.37\Rey^e\right)^{1/2}$ \cite{magnaudet1995accelerated, shi2021drag} and intersects with Eq.\,\eqref{eq:cri_vorE} at $\Rey^e\approx210$. This agrees with the critical Reynolds number for the onset of primary wake instability found for solid spheres \cite{citro2016linear}. In the following, we provide a more detailed discussion of the bifurcation sequence as $\mu^\ast$ increases for two selected values of $\Rey^e$: $\Rey^e=300$ and $\Rey^e=500$. These two cases are chosen as they exhibit a rich bifurcation sequence. 

\subsection{Bifurcation sequence at $\Rey^e=300$} \label{bif_seq1}
For the series of cases with $\Rey^e=300$, our simulation results indicate that as $\mu^\ast$ exceeds approximately 15, the axisymmetric flow first breaks down through a supercritical pitchfork bifurcation, leading to a transition into a steady planar-symmetric state. Then, as $\mu^\ast$ increases beyond approximately 30, a secondary Hopf bifurcation occurs, causing the flow to exhibit periodic oscillations while maintaining its planar symmetry. This latter flow behaviour persists up to the limit $\mu^\ast\to\infty$. 

\begin{figure*}[!htb]
\centerline{\includegraphics[scale=0.7]{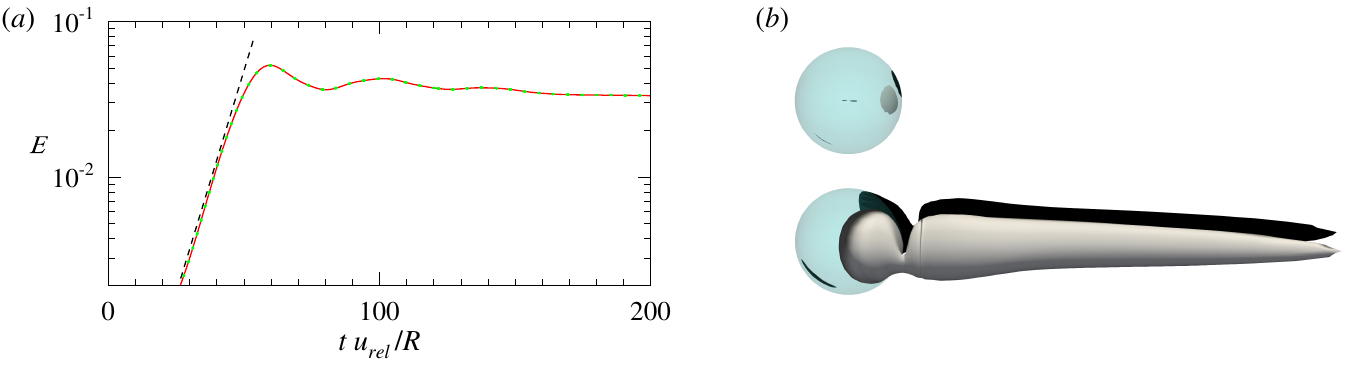}}
\caption{$(a)$ Time evolution of the total azimuthal energy $E$ \textcolor{black}{(solid red line) and the external azimuthal energy $E^e$ (dotted green line)} for $\mu^\ast=20$ and $\Rey^e=300$. The dashed line highlights the linear growth of $E$. $(b)$ Isosurfaces of the streamwise vorticity $\omega_x R/u_{rel}=\pm0.15$ inside (top) and outside (bottom) the droplet at steady state. Grey and black contours correspond to positive and negative values, respectively.}
\label{fig:e_vs_t}
\end{figure*}

\begin{figure*}[!htb]
\centerline{\includegraphics[scale=0.75]{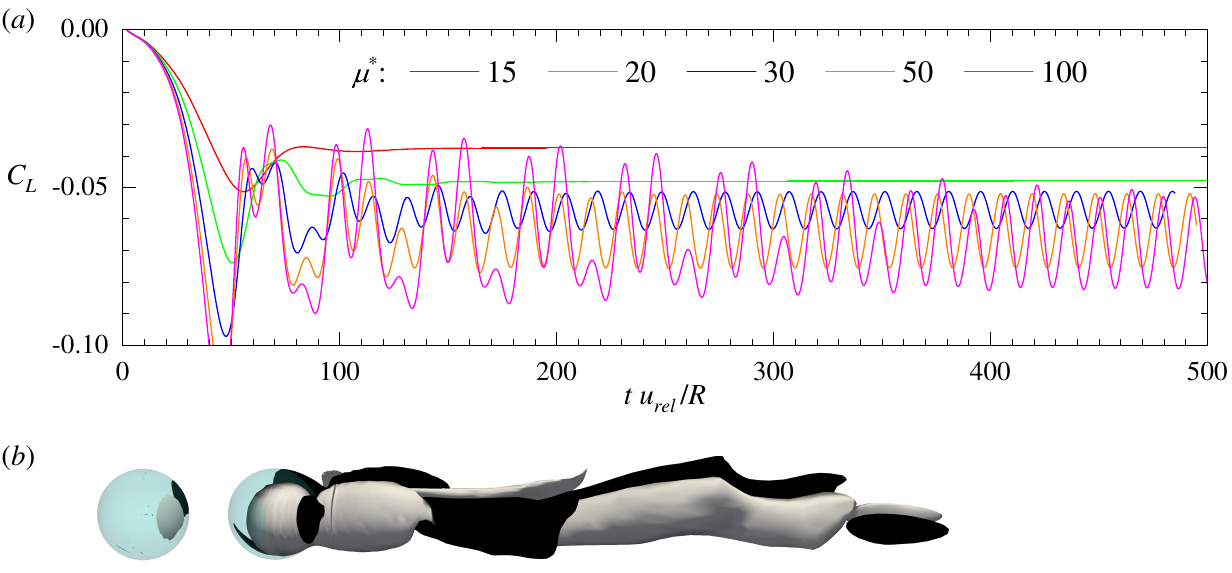}}
\caption{$(a)$ Time evolution of the lift coefficient $C_L$ for various $\mu^\ast$ at $\Rey^e=300$. $(b)$ Isosurfaces of the streamwise vorticity $\omega_x R/u_{rel}=\pm0.15$ inside (left) and outside (right) the droplet for $(\mu^\ast,\,\Rey^e)=(50,300)$.}
\label{fig:cl_vs_t}
\end{figure*}

The onset of axisymmetry breaking is characterised by an increase in the total azimuthal energy $E$. Figure \ref{fig:e_vs_t}$(a)$ shows the time evolution of $E$ \textcolor{black}{and $E^e$} for $\mu^\ast=20$. \textcolor{black}{The two time evolutions overlap, indicating a vanishingly small $E^i$ and that the internal flow remains axisymmetric throughout the evolution.} Symmetry breaking initiates at $t\,u_{rel}/R\approx20$, after which $E$ first exhibits a linear growth (highlighted by a dashed straight line), followed by a decrease in the growth rate. This behaviour clearly indicates that the bifurcation is supercritical \cite[p. 82]{strogatz2018nonlinear}. Figure \ref{fig:e_vs_t}$(b)$ depicts the distribution of streamwise vorticity $\omega_x$ in the fully developed state for $\mu^\ast=20$, which remains steady at this viscosity ratio. $\omega_x$ is negligibly small inside the droplet, indicating that the internal flow remains axisymmetric. In contrast, the external flow is strongly asymmetric, featured by a pair of counter-rotating streamwise vortices in the wake. This wake structure is similar to that associated with the primary wake instability of a solid sphere \cite{johnson1999flow, ghidersa2000breaking} and an oblate spheroidal bubble \cite{2007_Magnaudet, yang2007linear}, indicating that the axisymmetry breaking observed here is governed by an azimuthal mode with wavenumber $m=1$.  

The pair of counter-rotating vortices seen in figure \ref{fig:e_vs_t}$(b)$ is known to induce a transverse or lift force in the symmetry plane, whose orientation is selected by some initial disturbance. In our numerical setup, a weak streamwise linear shear flow of $10^{-4}\, y \left(u_{rel}/R\right) \,\boldsymbol{e}_x$ is imposed to trigger the bifurcation. \textcolor{black}{The velocity variation at the droplet scale is only $10^{-4}u_{rel}$, making it negligibly small compared to the main stream of the ambient flow.} The induced lift force points along $-\boldsymbol{e}_y$.
Figure \ref{fig:cl_vs_t}$(a)$ presents the time evolution of the lift coefficient $C_L$, defined as $F_L = C_L \pi R^2 \rho^e u_{rel}^2 / 2$, where $F_L$ is the component of the hydrodynamic force along $\boldsymbol{e}_y$, for $\mu^\ast$ ranging from 15 to 100. In the fully developed state, $C_L$ increases with increasing $\mu^\ast$ and becomes time-dependent once $\mu^\ast$ exceeds approximately 30. Beyond this value, the evolution of $C_L$ is nearly sinusoidal, suggesting that this force component responds almost linearly to wake dynamics.

Figure \ref{fig:cl_vs_t}$(b)$ shows the structure of the streamwise vorticity at $\mu^\ast=50$. While the internal flow remains axisymmetric, unsteadiness becomes prominent in the wake, with alternating positive and negative values of $\omega_x$ within each vortex thread. Indeed, by examining the time history of $C_L(t)$ shown in figure \ref{fig:cl_vs_t}$(a)$, we find that, regardless of $\mu^\ast$, the frequency $f_0$ associated with the oscillations of $C_L(t)$ corresponds to a Strouhal number $St = 2f_0 R/u_{rel}$ of 0.135. This value closely matches the Strouhal number obtained for both solid spheres \cite{johnson1999flow} and oblate spheroidal bubbles \cite{2007_Magnaudet} at the same external Reynolds number. This agreement strongly suggests that the secondary Hopf bifurcation is supercritical, as in the latter two cases.

\subsection{Bifurcation sequence at $\Rey^e=500$} \label{bif_seq2}

\begin{figure*}[!htb]
\centerline{\includegraphics[scale=0.75]{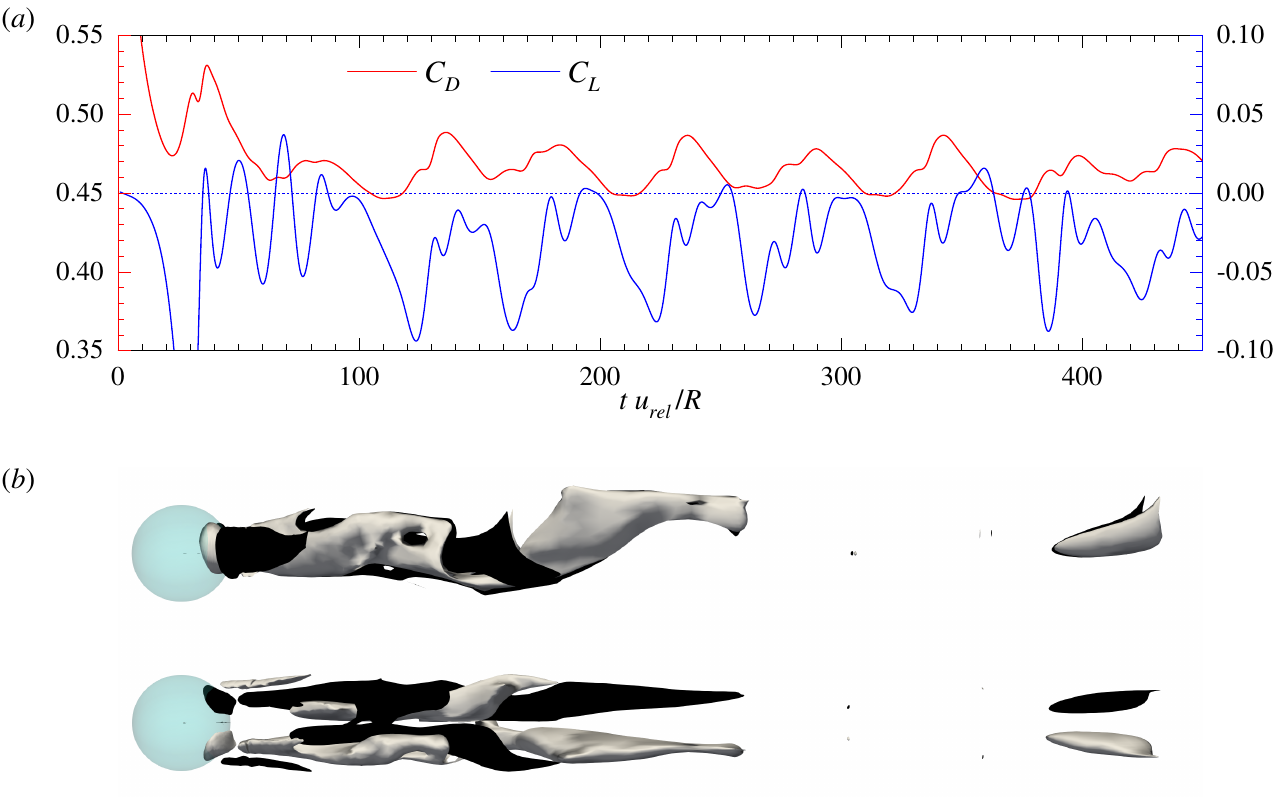}}
\caption{$(a)$ Time evolution of the drag and lift coefficients for $(\mu^\ast,\,\Rey^e)=(15,\,500)$. $(b)$ Isosurfaces of the streamwise vorticity $\omega_x R/u_{rel}=\pm0.15$ viewed from two perpendicular perspectives for the same case as in panel $(a)$.}
\label{fig:cd-cl_vs_t_vr15}
\end{figure*}

For the series of cases with $\Rey^e=500$, the primary bifurcation occurs at approximately $\mu^\ast=8$. The resulting flow is steady and has a structure similar to that observed for the case with $(\mu^\ast,\,\Rey^e)=(20, 300)$ (see figure \ref{fig:e_vs_t}$b$). As $\mu^\ast$ increases beyond approximately 12 but remains below 50, the post-bifurcation flow becomes unsteady while still preserving the symmetry with respect to the plane defined by the primary bifurcation. Figure \ref{fig:cd-cl_vs_t_vr15}$(b)$ shows the isocontours of the streamwise vorticity for $\mu^\ast=15$, viewed from two perpendicular directions. The wake structure exhibits reflectional symmetry, similar to that seen in figure \ref{fig:cl_vs_t}$(b)$. However, the vortical structures are more complex than those at $\Rey^e=300$, suggesting the potential presence of more than one unstable mode. To further examine this, figure \ref{fig:cd-cl_vs_t_vr15}$(a)$ shows the time evolution of the drag and lift coefficients for the present case. The variations of $C_D(t)$ and $C_L(t)$ reveal the presence of two distinct modes. Based on the time series starting from $t\,u_{rel}/R=100$, we identified a primary mode, characterised by small but rapid fluctuations, with a reduced frequency of $St = 2f_1R/u_{rel} \approx 0.154$. The secondary mode, exhibiting larger but slower fluctuations, has a reduced frequency of approximately $St = 2f_2R/u_{rel} \approx 0.04$.

\begin{figure*}[!htb]
\centerline{\includegraphics[scale=0.75]{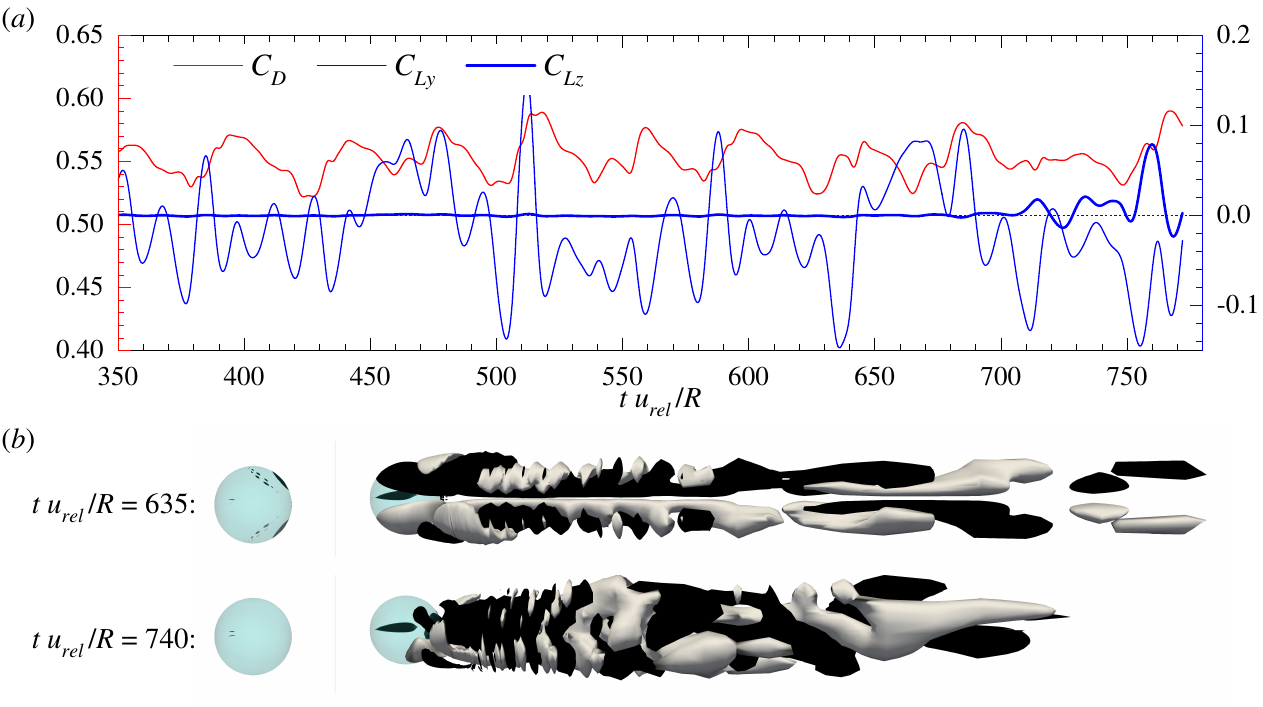}}
\caption{$(a)$ Time evolution of the drag and lift coefficients for $(\mu^\ast,\,\Rey^e)=(100,\,500)$. $(b)$ Isosurfaces of the streamwise vorticity $\omega_x R/u_{rel}=\pm0.25$ at two different instants. In both panels, $\boldsymbol{e}_y$ points inwards, and the flow moves from left to right.}
\label{fig:cd-cl_vs_t_vr100}
\end{figure*}

At $\mu^\ast=15$, the secondary mode remains relatively weak, such that $C_L$ remains mostly negative as it oscillates over time. As $\mu^\ast$ increases beyond 15, the secondary mode becomes more energetic, causing $C_L$ to occasionally reach positive values comparable to its negative maximum. Moreover, the planar symmetry is no longer preserved beyond a critical viscosity ratio of approximately 50. A representative case is $\mu^\ast=100$, for which the time evolution of the drag and lift coefficients is shown in figure \ref{fig:cd-cl_vs_t_vr100}$(a)$. Since the flow in the fully developed state is now fully three-dimensional, the lift force has two components: one, denoted as $C_{Ly}$, corresponds to the component in the symmetry plane determined by the primary bifurcation, while the other, denoted as $C_{Lz}$, is the projection of the transverse force along $\boldsymbol{e}_z$, where $\boldsymbol{e}_z \equiv \boldsymbol{e}_x \times \boldsymbol{e}_y$. 
It can be inferred that the flow remains planar symmetric up to approximately $t\,u_{rel}/R=700$, during which $C_{Lz}$ remains negligibly small, while $C_{Ly}$ and $C_D$ oscillate similarly to the case with $\mu^\ast=15$ (figure \ref{fig:cd-cl_vs_t_vr15}$a$). A similar evolution of the flow structure with increasing external Reynolds number, particularly the emergence and amplification of the secondary unstable mode prior to the transition to a fully three-dimensional flow regime (i.e. the loss of planar symmetry), has been reported in prior studies on spheroidal bubbles \cite{2007_Magnaudet} and solid spheres \cite{2000_Tomboulides}. 

While not shown in detail, the bifurcation sequence observed at higher $\Rey^e$ is similar to that at $\Rey^e=500$, except that the critical viscosity ratios marking the transitions between successive bifurcation states decrease as $\Rey^e$ increases. For instance, in the series of cases at $\Rey^e=700$, the flow transitions into a steady planar-symmetric state at $\mu^\ast=8$, then becomes unsteady for $\mu^\ast=10$, comprising two distinct unstable modes as in the case of $(\mu^\ast,\,\Rey^e)=(15,500)$, and ultimately transitions to a fully three-dimensional state (i.e. loss of planar symmetry) at $\mu^\ast=15$. Notably, in all these bifurcation sequences, the internal flow remains axisymmetric, meaning that flow asymmetry only develops in the external fluid, specifically in the wake downstream of the droplet.

\subsection{Influence of the internal Reynolds number} \label{inf_rei}
Thus far, we have examined the influence of the viscosity ratio and the external Reynolds number on the onset of various external bifurcations, as well as the structure of the three-dimensional flow field associated with these bifurcations. The internal Reynolds number, on the other hand, has been set at relatively large values, i.e. $\Rey^i\geq500$, and its influence has not yet been thoroughly explored. In this section, we investigate a series of cases at $\Rey^e=300$ to examine the effect of $\Rey^i$ over a wide range of $\mu^\ast$.
\begin{figure*}[!htb]
\centerline{\includegraphics[scale=0.75]{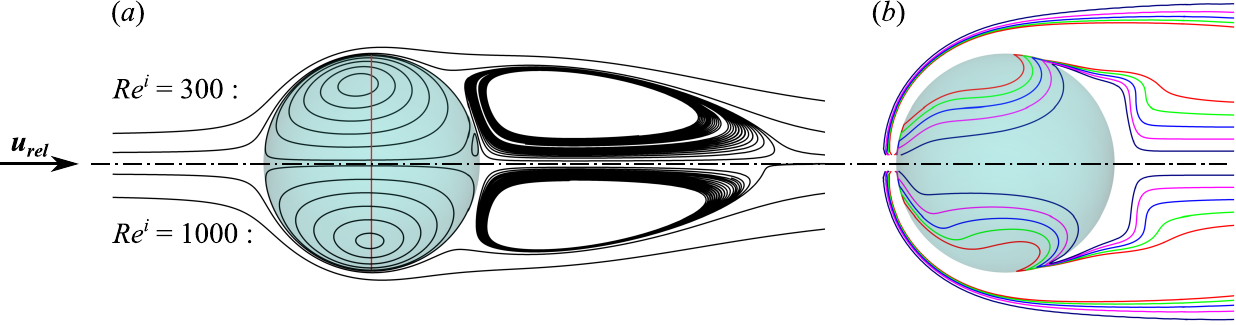}}
\caption{$(a)$ Streamlines and $(b)$ isocontours of the azimuthal vorticity $\omega_\phi$ around the droplet for $\Rey^i = 300$ (top panels) and $\Rey^i = 1000$ (bottom panels) with $(\mu^\ast,\Rey^e)=(10,300)$. In panel $(b)$, coloured lines represent $-\omega_\phi R/u_{rel} = 1$ (red), 0.8 (green), 0.6 (blue), 0.4 (magenta), and 0.2 (navy).}
\label{fig:axi_streamline_rei}
\end{figure*}

\begin{figure*}[!htb]
\centerline{\includegraphics[scale=0.75]{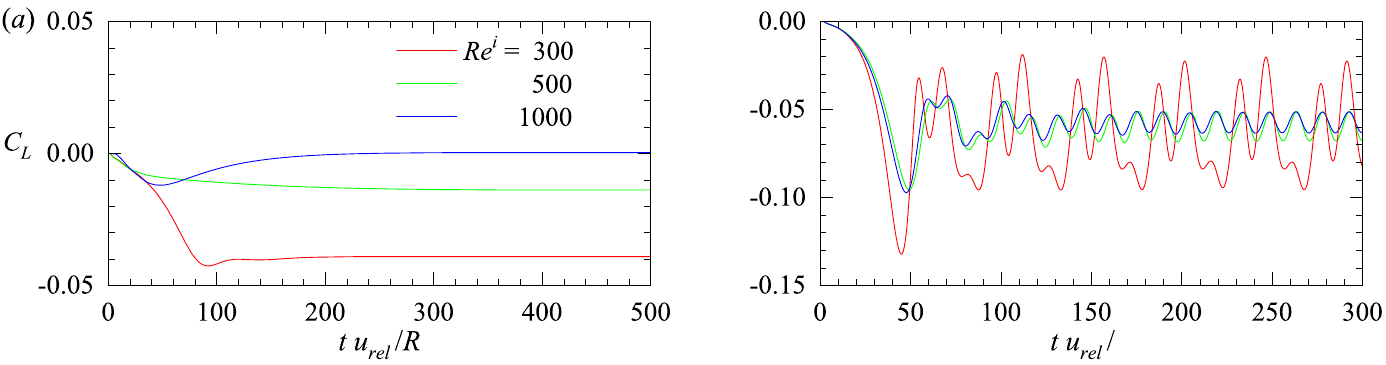}}
\caption{Time evolution of the lift coefficient at $\Rey^e=300$ for different $\Rey^i$. $(a)$ $\mu^\ast = 10$; $(b)$ $\mu^\ast = 30$.}
\label{fig:cl_vs_t_rei}
\end{figure*}

\begin{figure*}[!htb]
\centerline{\includegraphics[scale=0.75]{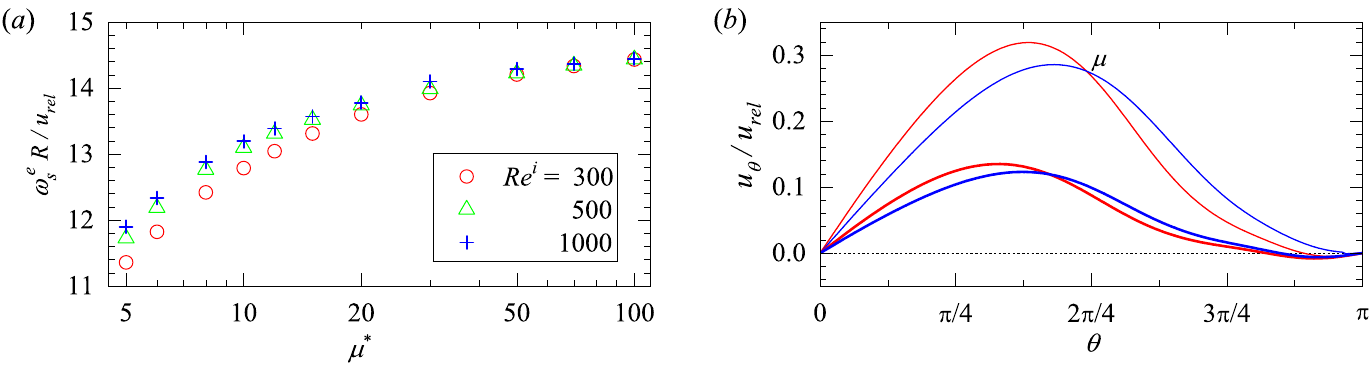}}
\caption{$(a)$ Maximum external surface vorticity $\omega_s^e$ as a function of $\mu^\ast$ for different $\Rey^i$. $(b)$ Tangential velocity $u_\theta$ at the droplet surface as a function of the meridional angle $\theta$ ($\theta=0$ and $\pi$ correspond to the front and rear stagnation points, respectively) for $\Rey^i=300$ (red lines) and $\Rey^i=1000$ (blue lines) at two different $\mu^\ast$.}
\label{fig:vor_e_rei}
\end{figure*}

Figure \ref{fig:axi_streamline_rei}$(a)$ compares the streamlines around and inside the droplet at $\Rey^i=300$ (top) and $\Rey^i=1000$ (bottom). In both cases, we set $(\mu^\ast,\,\Rey^e)=(10, 300)$ and impose axisymmetry on the entire flow. Inside the droplet, the streamlines exhibit stronger fore-aft asymmetry at the smaller $\Rey^i$, with a weak recirculation visible in the rear region. Outside the droplet, a strong recirculating flow is observed at the rear, which is slightly more pronounced at the smaller $\Rey^i$. Specifically, the reattachment length (measured from the rear stagnation point) is approximately $2.79R$ for $\Rey^i=300$, compared with $2.64R$ for $\Rey^i=1000$. Figure \ref{fig:axi_streamline_rei}$(b)$ presents the isocontours of the azimuthal vorticity for these two cases. The key difference lies in the tilting of the isocontours in the wake, which is clearly more pronounced at the smaller $\Rey^i$ (for instance, comparing the iso-lines at $-\omega_\phi R/u_{rel} = 0.6$ in both cases). This distinction is significant, as following the physical mechanism of external bifurcation outlined in \S\,\ref{axi_wak}, a stronger tilting of the $\omega_\phi$-isocontours suggests an earlier onset of the external bifurcation. Indeed, results from simulations performed in a fully three-dimensional domain indicate that, in these two cases, the axisymmetry of the flow already breaks down at $\Rey^i=300$, whereas it remains stable at $\Rey^i=1000$.

Figure \ref{fig:cl_vs_t_rei}$(a)$ shows the time evolution of the lift coefficient obtained in a fully three-dimensional configuration at three different $\Rey^i$ for $(\mu^\ast,\,\Rey^e)=(10, 300)$. At $\Rey^i=1000$, $C_L$ in the final state is negligibly small, indicating that the axisymmetry of the flow is preserved. In contrast, for the two smaller $\Rey^i$ values, $C_L$ becomes non-zero, confirming that axisymmetry breaking occurs. Figure \ref{fig:cl_vs_t_rei}$(b)$ presents the same results as panel $(a)$ but for $\mu^\ast=30$. At this viscosity ratio, axisymmetry breaks down, and vortex shedding occurs for all three considered $\Rey^i$. The time-averaged $C_L$ in the fully developed state is found to be $-0.057$, $-0.060$, and $-0.065$ for $\Rey^i=1000$, 500, and 300, respectively, revealing a gradual increase in lift with decreasing $\Rey^i$. Additionally, at $\Rey^i=300$, the evolution of $C_L$ suggests the presence of a secondary mode with a Strouhal number approximately three times higher than the primary mode, whereas this secondary mode remains inactive at the other two $\Rey^i$ values.

To understand why smaller $\Rey^i$ promotes external bifurcation in cases with fixed $(\mu^\ast,\Rey^e)$, we examine the maximum external surface vorticity $\omega_s^e$ as a function of $\mu^\ast$ for different $\Rey^i$. As seen in figure \ref{fig:vor_e_rei}$(a)$, $\omega_s^e$ is slightly lower at smaller $\Rey^i$, particularly when $\mu^\ast < 20$. Notably, at $\mu^\ast=10$, $\omega_s^e$ is only $12.4u_{rel}/R$, approximately $90\%$ of the threshold (around $14u_{rel}/R$) predicted by Eq.\,\eqref{eq:cri_vorE}. Thus, Eq.\,\eqref{eq:cri_vorE} provides only a sufficient condition for external bifurcation. For droplets with $\Rey^i<500$, the actual threshold $\omega_s^e$ may generally be lower than this prediction. Exploring the dependence of the threshold $\omega_s^e$ on $\Rey^i$ is computationally expensive. Instead, we provide insight into why smaller $\Rey^i$ promotes external bifurcation. We find this promotion to be closely linked with the immobilisation of the rear part of the droplet. 
To elaborate, figure \ref{fig:vor_e_rei}$(b)$ compares the tangential velocity $u_\theta$ at the droplet surface as a function of the meridional angle $\theta$ ($\theta=0$ and $\pi$ correspond to the front and rear stagnation points, respectively) for $\Rey^i=300$ (red lines) and $\Rey^i=1000$ (blue lines) at two different $\mu^\ast$. In both cases, the tangential velocity is larger at the smaller $\Rey^i$ for $\theta < \theta_c$ (with $\theta_c\approx \pi/2$ and $3\pi/8$ for $\mu^\ast=10$ and 30, respectively), whereas the opposite is true for $\theta > \theta_c$. In other words, decreasing $\Rey^i$ increases surface mobility in the front part of the droplet while reducing it in the rear. Since the onset of wake instability originates from disturbances growing in the core of the standing eddy \cite{chomaz2005global}, which is located closer to the rear part, we strongly suspect that this reduction in mobility promotes the onset of the external bifurcation at smaller $\Rey^i$. Of course, what we propose here is merely a plausible scenario based on our numerical observations. For a rigorous justification, a stability analysis of the flow in the transition region would be required in future work.

\section{Conclusion} \label{conclusion}
We carried out three-dimensional numerical simulations of a uniform flow past a fixed spherical droplet over a wide range of dimensionless numbers, namely, the viscosity ratio $\mu^\ast$, the external Reynolds number $\Rey^e$, and the internal Reynolds number $\Rey^i$. We focused on highly viscous droplets where $\mu^\ast\geq5$ and explored the parameter regime in which the axisymmetry of the flow breaks down due to an external flow bifurcation. By examining the vortical structure in cases close to the bifurcation threshold, we found that the onset of bifurcation is closely related to the tilting of the azimuthal vorticity $\omega_\phi$ in the wake. Specifically, the bifurcation sets in once the isocontours align almost perpendicularly to the symmetry axis. This behaviour resembles that observed in the wake of a spheroidal bubble close to wake instability \cite{2007_Magnaudet}, suggesting that the underlying mechanism of symmetry breaking is essentially the same as that observed for particles and oblate spheroidal bubbles, as elaborated in \cite{2007_Magnaudet}. We showed that the critical curve delimiting the unstable region may be recast in terms of the maximum vorticity on the external side of the droplet surface, $\omega_s$, yielding an empirical criterion similar to that of \cite{2007_Magnaudet} to determine whether the axisymmetric flow is stable. This criterion was established based on data at high $\Rey^i$ and applies over wide ranges of $\mu^\ast$ and $\Rey^e$, provided that $\Rey^i\geq500$.

Then, we selected two particular external Reynolds numbers, $\Rey^e=300$ and $\Rey^e=500$, to study the flow evolution with increasing viscosity ratio $\mu^\ast$. For both series of cases, $\Rey^i$ was kept fixed at 1000. In the series of cases with $\Rey^e = 300$, the flow first undergoes a supercritical regular bifurcation, yielding a steady non-axisymmetric flow with planar symmetry. In this configuration, the wake exhibits a pair of counter-rotating streamwise vortices, yielding a transverse or lift force in the symmetry plane. As the viscosity ratio increases further, a secondary supercritical Hopf bifurcation occurs. The sign of the streamwise vorticity shed in each vortex thread then changes periodically with a frequency that remains independent of the viscosity ratio as $\mu^\ast\to\infty$. Planar symmetry is still maintained in this regime, as well as the sign of the mean lift force. 
For the series of cases with $\Rey^e = 500$, the first two bifurcations are similar to those observed at $\Rey^e = 300$, but they reveal a progressively increasing complexity in wake dynamics. At sufficiently large $\mu^\ast$, the wake even loses the planar symmetry imposed by the primary bifurcation. While first documented in the present work, this bifurcation sequence, particularly the first two bifurcations, closely resembles those observed in previous studies for fixed rigid spheres, disks, and oblate spheroidal bubbles \cite{2012_Ern}. This suggests that the sequence is a characteristic feature of axisymmetric bluff bodies, irrespective of their precise shape or surface nature. This also indicates that the mechanism driving the external flow bifurcation depends primarily on the amount of vorticity generated on the external side of the body surface, rather than the specific boundary condition imposed on it.

\textcolor{black}{Despite the similarities discussed above, droplets exhibit distinct characteristics due to the internal flow, which does not play an active role in triggering external flow bifurcation for oblate spheroidal bubbles and solid particles. Given the subtle role played by the internal flow in droplets (see the discussion in \S\,\ref{inf_rei} for details),} a universal criterion determining the onset of the external bifurcation based solely on the maximum external surface vorticity, $\omega_s^e$, appears to be feasible only when the internal Reynolds number is high, i.e. for $\Rey^i=\mathcal{O}(1000)$. At moderate $\Rey^i$, our numerical results indicate that external bifurcation sets in at values of $\omega_s^e$ lower than the critical threshold predicted by Eq.\,\eqref{eq:cri_vorE}. A detailed examination of the flow fields suggests that decreasing $\Rey^i$ reduces surface mobility at the rear of the droplet, thereby promoting external bifurcation. To gain a better understanding of the role of $\Rey^i$, future work employing either DNS or linear stability analysis at values of $\Rey^i$ near the onset of the external bifurcation are necessary. Such an investigation would also facilitate extending the criterion for determining the onset of the external bifurcation to flow regimes with moderate $\Rey^i$.

\end{multicols}
\begin{multicols}{2}

\Acknowledgements{This work was supported by the Deutsche Forschungsgemeinschaft (DFG, German Research Foundation) through grant number 501298479.}

\bibliographystyle{spmpsci}
\bibliography{AMS-template-2022}

\end{multicols}

\makeentitle

\end{document}